\documentclass[10pt,preprint2]{aastex}
\shorttitle{Extrasolar Moons}
\shortauthors{Barnes \& O'Brien}

\begin{document}

\title{Stability of Satellites Around Close-in Extrasolar Giant Planets}

\author{Jason W. Barnes and D. P. O'Brien}
\affil{Department of Planetary Sciences, University of Arizona, Tucson, AZ, 85721}
\email{jbarnes@c3po.lpl.arizona.edu, obrien@lpl.arizona.edu}

\newpage

\begin{abstract} 

We investigate the long-term dynamical stability of hypothetical moons orbiting
extrasolar giant planets.  Stellar tides brake a planet's rotation and, together
with tidal migration, act to remove satellites;  this process limits the
lifetimes of larger moons in extrasolar planetary systems.  Because more massive
satellites are removed more quickly than less massive ones, we are able to derive an
upper mass limit for those satellites that might have survived to the present day. 
For example, we estimate that no primordial satellites with masses greater than
$7\times10^{-7}~\mathrm{M_\oplus}$ ($\sim \hspace{-3pt}70~\mathrm{km}$ radius for
$\rho=3~\mathrm{g~cm^{-3}}$) could have survived around the transiting planet HD209458b
for the age of the system.  No meaningful mass limits can be placed on moons
orbiting Jovian planets more than $\sim0.6~\mathrm{AU}$ from their parent stars. 
Earth-like moons of Jovian planets could exist for $5~\mathrm{Gyr}$ in systems where
the stellar mass is greater than $0.15~\mathrm{M_\odot}$.    Transits show the most
promise for the discovery of extrasolar moons --- we discuss prospects for satellite
detection via transits using space-based photometric surveys and the limits on the
planetary tidal dissipation factor $Q_p$ that a discovery would imply.

\end{abstract}


\section{INTRODUCTION} 

Each of the giant planets in our solar system posesses a satellite system. Since the
discovery of planets in other solar systems \citep{2000prpl.conf.1285M}, the
question of whether these extrasolar planets also have satellites has become
relevant and addressible.  Extrasolar planets cannot be observed directly with
current technology, and observing moons around them poses an even greater
technical challenge.  However, high precision photometry of stars during transits of
planets can detect extrasolar moons either by direct satellite transit or through
perturbations in the timing of the planet transit \citep{1999A&AS..134..553S}. 
Using these techniques, \citet{2001ApJ...552..699B} placed upper limits of $1.2$
Earth radii ($~\mathrm{R_\oplus}$) and $3$ Earth masses  ($~\mathrm{M_\oplus}$) on
any satellites orbiting the transiting planet HD209458b based on the \emph{Hubble
Space Telescope} transit lightcurve.

The tidal bulge that a satellite induces on its parent planet perturbs the
satellite's orbit \citep[e.g.,][]{1986sats.book..117B}, causing migrations in
semimajor axis that can lead to the loss of the satellite.  For an isolated planet,
satellite removal occurs either through increase in the satellite's orbital
semimajor axis until it escapes, or by inward spiral until it impacts the planet's
surface \citep{1973ApJ...180..307C}.  In the presence of the parent star,
stellar-induced tidal friction slows the planet's rotation, and the resulting
planet-satellite tides cause the satellite to spiral inward towards the planet
\citep{1973MNRAS.164...21W,1973NaPS..242....23}.  This effect is especially
important for a planet in close proximity to its star, and has been suggested as the
reason for the lack of satellites around Mercury
\citep{1973MNRAS.164...21W,1973NaPS..242....23}.  

In this paper, we apply tidal theory and the results of numerical orbital
integrations to the issue of satellites orbiting close-in extrasolar giant planets.
We place limits on the masses of satellites that extrasolar planets may posess,
discuss the implications these limits have for the detection of extrasolar
satellites, and apply our results to the issue of Earth-like satellites orbiting
extrasolar giant planets.

\section{TIDAL THEORY AND METHODS}

According to conventional tidal theory, the relative values of the planetary
rotation rate, $\Omega_p$, and the orbital mean motion of the moon, $n_m$ (both in
units of $\mathrm{radians}/\mathrm{sec}$), determine the direction of orbital
evolution \citep[see e.g.,][]{SolarSystemDynamics}.  For a moon orbiting a planet
slower than the planet rotates ($n_m<\Omega_p$), the tidal bulge induced on
the planet by the satellite will be dragged ahead of the satellite by an angle
$\delta$, with $\tan(2\delta) = 1/Q_p$.  Here, $Q_p$ is the parameter describing
tidal dissipation within the planet \citep[after][]{1966Icar....5..375G}, with 
$1/Q_p$ equal to the fraction of tidal energy dissipated during each tidal cycle. 
Gravitational interactions between the tidal bulge and the satellite induce
torques that transfer angular momentum and dissipate energy, slowing the planet's
rotation and increasing the orbital semimajor axis of the satellite.  Conversely,
for satellites orbiting faster than their planet's rotation ($n_m>\Omega_p$), the
planet is spun up and the satellite's semimajor axis decreases.  The same
mechanism causes torques on the planet from its parent star which slow the
planet's rotation \citep{SolarSystemDynamics}.

The torque on the planet due to the tidal bulge raised by the moon ($\tau_{p-m}$) is
\citep{SolarSystemDynamics}
\begin{equation}
\label{eq:torquepm}
\tau_{p-m}~=~-\frac{3}{2} \frac{k_{2p}~G~M_m^2~R_p^5}{Q_p~a_m^6}
~\mathrm{sign}(\Omega_p-n_m) ~\mathrm{,}
\end{equation}
where $k_{2p}$ is the tidal Love number of the planet, $R_p$ is the radius of
the planet, and $G$ is the gravitational constant.  The term 
$\mathrm{sign}(\Omega-n_m)$ is equal to $1$ if $(\Omega-n_m)$ is
positive, and is equal to $-1$ if it is negative.  We obtain the expression for the
stellar torque on the planet by replacing $M_m$, the mass of the moon,
with $M_*$, the mass of the star; by replacing $a_m$, the semimajor axis of the
satellite's orbit with $a_p$, the semimajor axis of the planet's orbit about the
star (circular orbits are assumed); and by using the planet's mean orbital motion
$n_p$ instead of $n_m$:  
\begin{equation}
\tau_{p-*}~=~-\frac{3}{2} \frac{k_{2p}~G~M_*^2~R_p^5}{Q_p~a_p^6}
~\mathrm{sign}(\Omega_p-n_p)~\mathrm{.}
\end{equation}
The moon's semimajor axis, $a_m$, and the moon's mean motion, $n_m$, are related by
Kepler's law, $n_m^2 a_P^3=GM_p$.  These torques affect both $n_m$ and $\Omega_p$.  The rate of change of $\Omega_p$
is obtained by dividing the total torque on the planet by the planet's moment of
inertia:
\begin{equation}\label{eq:domegadt}
\frac{d\Omega_p}{dt}~=~\frac{\tau_{p-m}+\tau_{p-*}}{I_p}~\mathrm{,}
\end{equation}
where $I_p$ is the planet's moment of inertia. 
 
Under the circumstances studied in this paper, where a planet is orbited by a
much  smaller satellite, $\tau_{p-*}$ is much greater than $\tau_{p-m}$ for
most of the system's lifetime.  Because the moon's orbital moment of inertia
depends on $n_m$, the equivalent expression for $n_m$ is less trivial to derive. 
We obtain it by setting the torque equal to the rate of change of the angular
momentum and solving for $dn/dt$ using the planet's mass, $M_p$
\citep[e.g.,][]{1988Icar...74..153P}:
\begin{equation}\label{eq:dndt}
\frac{dn_m}{dt}~=~\frac{3~\tau_{p-m}~n_m^{4/3}}{M_m~(GM_p)^{2/3}}
~\mathrm{.}
\end{equation}
Given appropriate initial and boundary conditions, integration of Equations
\ref{eq:domegadt} and \ref{eq:dndt} 
determines the state of the system at any given time.

An important boundary condition for such an integration is the critical semimajor
axis, or the location of the outermost satellite orbit that remains bound to the
planet. This location must be within the planet's gravitational influence, or Hill
sphere, and has been generally thought to lie between $1/3$ and $1/2$ the radius of
the Hill sphere ($R_H$) \citep{1986sats.book..117B}, where
\begin{equation}
R_H=a_p\left(\frac{M_p}{3M_*}\right)^{1/3} ~\mathrm{.}
\end{equation} 
Recently,
\citet{1999AJ....117..621H} investigated the stability of planets in binary
star systems and their results are applicable to the planet-satellite
situation as well.  Through numerical integrations of a test
particle orbiting one component of a binary star system,
\citet{1999AJ....117..621H} found that for high mass ratio binaries, the
critical semimajor axis for objects orbiting the secondary in its orbital plane
is equal to a constant fraction ($f$) of the secondary's Hill radius, or
\begin{equation}\label{eq:acrit}
a_{crit} = f R_H ~\mathrm{.}
\end{equation}
We treat a star orbited
by a much less massive planet as a high mass ratio binary system and deduce that
the critical semimajor axis for a satellite orbiting the planet is $.36R_H$
($f=.36$) for 
prograde satellites \citep[from][Figure 1]{1999AJ....117..621H}.
This agrees closely with
\citet{1986sats.book..117B}. In fact, none of the prograde moons of
our solar system orbit outside this radius (see Table
\ref{table:hillsemimajoraxes}). \citet{1999AJ....117..621H} did not treat
objects in retrograde orbits (which are expected to be more stable than
prograde ones), so to treat possible captured satellites we take
$f_{retrograde}=0.50$ based on the solar system values for ${a_m}/{R_H}$ in
Table \ref{table:hillsemimajoraxes}.

\section{CONSTRAINTS ON SATELLITE MASSES}

Satellites orbiting close-in giant planets fall into one of three categories based
on the history of their orbital evolution.  Satellites that either start inside the
planet's synchronous radius (the distance from the planet where $n_m=\Omega_p$) or
become subsynchronous early in their lifetimes, as a result of the slowing
planetary rotation, spend their lives spiraling inward toward the cloud tops. 
Eventually these moons collide with the planet or are broken up once they migrate
inside the Roche limit.  Moons that start and remain exterior to the synchronous
radius evolve outward over the course of their lives and, given enough time, would
be lost to interplanetary space as a result of orbtial instability.  In between
these two is a third class of orbital history.  In this case, a satellite starts
well outside the synchronous radius and initially spirals outward, but its
migration direction is reversed when the planet's rotation slows enough to move
the synchronous radius outside the moon's orbit.  These moons eventually impact
the planet.

\begin{figure} 
\plotone{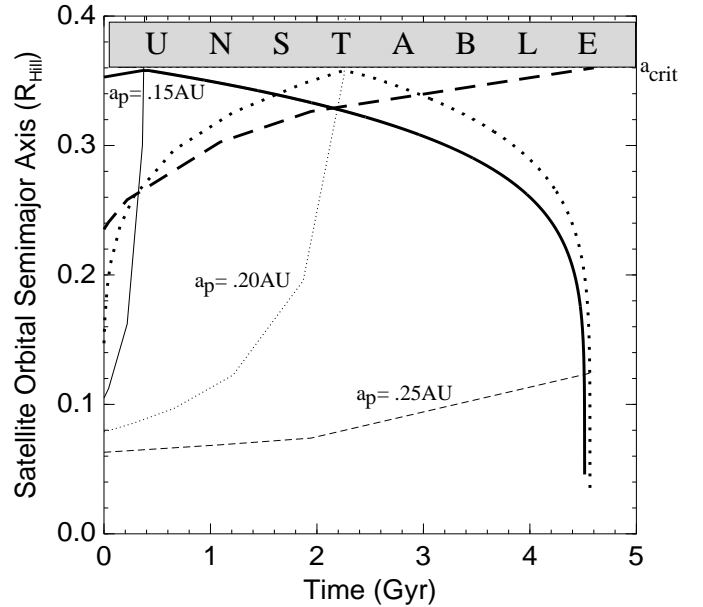}
\caption{Satellite orbital semimajor axis versus time for the
maximum-mass moon in three different hypothetical $4.6~\mathrm{Gyr}$ old 
$1~\mathrm{M_\odot}$,  $1~\mathrm{M_{Jup}}$ planetary systems.  The solid lines
represent a system with a planet-star distance of $0.15~\mathrm{AU}$, with the 
thick and thin lines corresponding to the satellite semimajor axis and planet
synchronous radius respectively.  Tides between this planet and its star spin
down the planet in short order, and the moon spends the majority of its lifetime
evolving inward through tidal interactions with the planet.  It is destroyed upon 
reaching the cloud tops of the planet.  The dotted lines correspond to 
a system in which the planet-star separation is $0.20~\mathrm{AU}$; this planet
is despun in just under half the age of the system.  Thus its maximum-mass moon
initially moves outward due to tidal influences, but later reverses direction due
to the spindown of the planet and eventually crashes into it.  This moon reverses 
direction at the critical semimajor axis (the outermost stable orbit point) because
doing so maximizes its orbital lifetime.  The dashed 
lines are for a planet orbital
semimajor axis of $0.25~\mathrm{AU}$.  The star's tidal torques on the planet
have less influence at this distance, and the planet does not despin sufficiently
over its lifetime to reverse the orbital migration of its maximum-mass satellite.
This satellite is lost into interplanetary space due to orbital instabilities.
\label{figure:avst}} 
\end{figure}

In order to determine which satellites might still exist around any given planet, we
determine the maximum lifetime for a moon with a given mass in each orbital
evolution category.  Inward-evolving satellites should maximize their lifetime by
starting as far from the planet as possible, at the critical semimajor axis
$a_{crit}$ (Equation \ref{eq:acrit}), and spiraling inward all the way to the
planet.  Outward-evolving satellites can survive the longest if they start just
outside the synchronous radius of the planet, then spiral outward to the critical
semimajor axis.  The maximum lifetime for the out-then-in case occurs when a
satellite reverses migration direction at the outermost possible point, the critical
semimajor axis.  In this case, the moon starts at the semimajor axis that allows for
it to have reached $a_{crit}$ by the time its planet's synchronous radius also
reaches $a_{crit}$, thus maximizing the time for its inward spiral (see Figure
\ref{figure:avst}).  For a given satellite mass, if the maximum possible lifetime is
shorter than the age of the system, then such a satellite could not have survived to
the present.  Because the orbits of  higher-mass satellites evolve more quickly than
those of lower-mass satellites (Equation \ref{eq:dndt}), an upper limit can be
placed on the masses of satellites that could still exist around any given planet.

\subsection{Analytical Treatment}

\begin{figure} 
\plotone{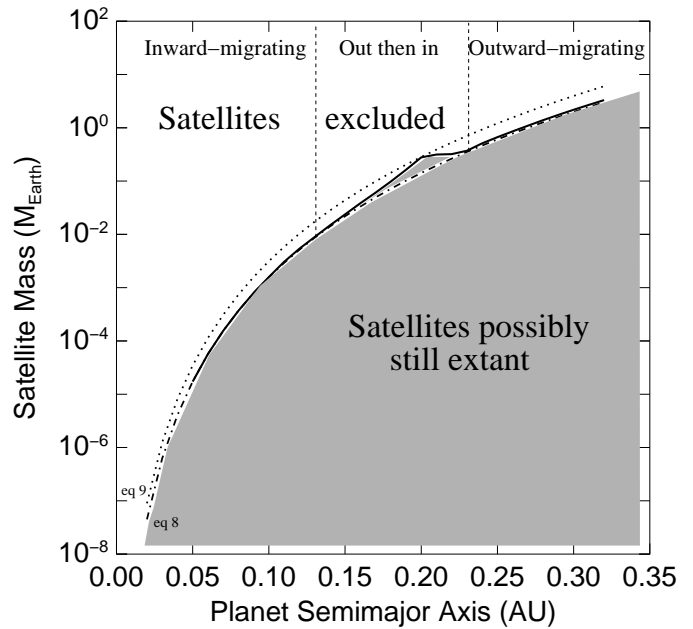}
\caption{Stability diagram for satellites in a hypothetical $4.6~\mathrm{Gyr}$
old $1~\mathrm{M_\odot}$,  $1~\mathrm{M_{Jup}}$ planetary system as a
function of the planet's orbital semimajor axis.  The solid line represents the
results from numerical integrations of Equations \ref{eq:domegadt} and
\ref{eq:dndt}, and the broken lines are the analytical approximations given in
Equations \ref{eq:Rhilltransverse} (lower, dot-dashed) and \ref{eq:Rhilltransverse2}
(upper, dotted).  Satellites above this line are excluded, while those which
lie below the line may or may not still exist depending on their specific orbital
evolutionary histories.  For this specific case, below $a_p\sim0.15\mathrm{AU}$ satellite masses are
limited by their inward migration, above $a_p\sim0.23\mathrm{AU}$ by their outward
migration, and in between by outward followed by inward migration with the
reversal being the result of slowing planetary rotation.
\label{figure:Mvsa}} 
\end{figure}

For a given semimajor axis of a moon, $a_m$, the migration rate is the same whether
the moon is moving inward or outward (with the assumption that $Q_p$ is independent
of the tidal forcing frequency $\Omega_p-n_m$), and the migration rate is much
faster for satellites close to their parent planets.  For both the inward- and
outward-migrating categories, the total lifetime of a satellite ($T$) is
well-characterized by the time necessary for a satellite orbit to traverse the
entire region between the critical semimajor axis ($a_m=a_{crit}$) and the planet's
surface ($a_m=R_p$), \citep{SolarSystemDynamics}: 
\begin{equation} 
\label{eq:traversaltime} T =
\frac{2}{13} \bigg(a_{crit}^{13/2} - R_{p}^{13/2}\bigg) \frac{Q_{p}}{3 k_{2{p}} M_m
R_p^{5}} \sqrt{\frac{M_p}{G}} ~\mathrm{.} 
\end{equation} 
Since $R_p \ll a_{crit}$
and the exponents are large, the $R_p$ term inside the parenthesis can be
neglected.  Substituting $a_{crit}=f R_H$ (Equation \ref{eq:acrit}), allowing $T$ to be equal to the age of
the system, and solving for $M_m$ collectively  result in an analytical expression
for the maximum possible extant satellite mass in both the inward and outward cases,
\begin{equation} 
\label{eq:Rhilltransverse} 
M_m \le \frac{2}{13} \bigg( \frac{(fa_p)^3}{3M_*} \bigg)^{13/6} \frac{M_p^{8/3}~Q_p}{3k_{2p}TR_p^5\sqrt{G}}~\mathrm{,}
\end{equation} 
which is the equation of the bottom, dot-dashed line in Figure
\ref{figure:Mvsa}.  In the case of satellites that evolve outward then inward, the
spindown of the planet is important.  These moons can be saved temporarily by the
reversal of their orbital migration.  This reversal prolongs their lifetimes, but by
less than a factor of two, because the satellite transverses the region where $a_m <
a_{crit}$ twice.  The upper mass limit for these satellites is 
\begin{equation}
\label{eq:Rhilltransverse2} 
M_m \le \frac{4}{13} \bigg( \frac{(f a_p)^3}{3M_*} \bigg)^{13/6} \frac{M_p^{8/3}~Q_p}{3k_{2p}TR_p^5\sqrt{G}}~\mathrm{,} 
\end{equation}
and this limit is plotted as the upper dotted line in Figure \ref{figure:Mvsa}.  

We obtain the boundaries between the in, out-then-in, and out cases by
comparing the time necessary to despin the planet to the age of the system. 
The time necessary to spin down the planet to the point that the synchronous
radius becomes exterior to the critical semimajor axis is
equal to \citep{1996ApJ...459L..35G}
\begin{equation} \label{eq:Tspindown}
T_{spindown}=Q_p (\Omega_{p0} - \Omega_{p1})
\left( \frac{R_p^3 M_p}{GM_*^2} \right) ~\mathrm{,}
\end{equation}
where $\Omega_{p0}$ is the initial planetary rotation rate and $\Omega_{p1}$ is the
rotation rate at which the planet's synchronous radius is coincident with
$a_{crit}$.  At this point, from Kepler's law we infer that
\begin{equation} \label{eq:omegap1}
\Omega_{p1} = n_{crit} = \left( \frac{G M_p}{a_{crit}^2}\right)^{(1/3)} ~\mathrm{.}
\end{equation}
For the case where $T_{spindown} \ll T$ (where $T$ is
the system liftime), the
maximum-mass moon evolves inward and Equation \ref{eq:Rhilltransverse} should
be used.  When the system age is greater than this spindown time
($T_{spindown} \gg T$), moons evolve outward and again Equation
\ref{eq:Rhilltransverse} is valid.  However, when $T_{spindown} \sim
T$, the reversal of satellite orbital migration is important and
Equation \ref{eq:Rhilltransverse2} provides a more robust upper mass limit for
surviving satellites.

These results for the maximum $M_m$ are limited by the requirement that the rate of
angular momentum transfer between the planet and the satellite must be less than
that between the planet and the star when $a_m$ is relatively large (i.e.,
$\tau_{p-*} > \tau_{p-m}$ in Equation \ref{eq:domegadt}) such that synchronization
between the planet and moon does not occur.  In the case of rocky satellites
orbiting gaseous planets, this condition is met.  For large moon-planet mass ratios
or large $a_p$, this assumption breaks down, yielding a situation more closely
resembling the isolated planet-satellite systems treated in
\citet{1973ApJ...180..307C}.  In this case the planet and moon can become locked
into a 1:1 spin-orbit resonance with each another, halting the satellite's orbital
migration and extending its lifetime.  For extrasolar Jovian planets
($0.3\mathrm{M_\odot}<M_p<13.0\mathrm{M_\odot}$) this occurs
when satellite masses become very large, i.e. greater than $8\mathrm{M_\oplus}$ for
a $1\mathrm{M_{Jup}}$ planet.  Such a moon is large enough to accrete hydrogen gas
during its formation, however, and in such a case the system is better treated as a
binary planet, taking into account the tidal torques of each body on the mutual
orbit.  We do not address that situation here.

We assume prograde, primordial satellites, but objects captured into
orbit by a planet late in its life could also remain in orbit.  We do not treat
the physics of satellite capture, but the lifetimes of such moons would be
affected by the same processes described above if prograde, and limited by
inward migration like Neptune's moon Triton \citep{1966AJ.....71..585M} if
retrograde.  In the case of retrograde, captured moons, the following upper
limit on their survival lifetime can be placed by rearranging Equation
\ref{eq:Rhilltransverse}:
\begin{equation}
\label{eq:retrotransverse}  
T_{max} = \frac{2}{13} \bigg( \frac{(f_R a_p)^3}{3M_*} \bigg)^{13/6} 
\frac{M_p^{8/3}~Q_p}{3k_{2p}M_mR_p^5\sqrt{G}}~\mathrm{.}
\end{equation}

Our analysis assumes a single satellite system.  Inward-migrating moons could
not be slowed significantly by entering into a resonance with another satellite
further in because the interior satellite would be migrating faster (due to
the $a_m^{-6}$ dependance of the torque in Equation \ref{eq:torquepm}), unless
its mass is less than $.08~M_m$ (assuming a 2:1 resonance).  Slowly-migrating
moons exterior to the satellite in question cannot slow its orbital migration
because objects in diverging orbits cannot be captured into resonances. 
However, outward-migrating satellites could have their lifetimes extended by
entering into a resonance with an exterior neighbor through intersatellite
angular momentum transfer \citep{1965MNRAS.130..159G}, similar to the resonances
currently slowing the outward migration of Io from Jupiter.  Thus some
outward-limited satellites above the mass limit derived in Equation
\ref{eq:Rhilltransverse} may still survive because of resonances entered into
earlier in their lifetimes.

\newpage

\subsection{Numerical Treatment}

To verify the limits stipulated in Equations \ref{eq:Rhilltransverse} and
\ref{eq:Rhilltransverse2}, we integrate Equations \ref{eq:domegadt} and
\ref{eq:dndt} numerically, from the initial rotation rate and semimajor axis until
the satellite's demise either through impact with the planet or through orbital
escape.  We use an adaptive stepsize Runge-Kutta integrator from
\citet{1992nrca.book.....P} and have verified that it reproduces our analytical
results for small satellite masses.  We also assume that only $\Omega_p$ and $a_m$
change over time --- other planetary parameters such as $Q_p$, $a_p$, $R_p$, and
all others are taken to be constant for the length of the integration.  The
expected changes in the planet's orbital semimajor axis $a_p$ over the course of
the integration do not significantly affect the calculations, and larger planetary
radii $R_p$ in the past would only serve to further reduce lifetime of a given
satellite beyond what we have calculated here, pushing the upper surviving
satellite mass lower.

For each planet, we determine the maximum satellite mass that could survive for the
observed lifetime of the system by optimizing the initial semimajor axis of the
satellite so as to maximize its lifetime and then tuning the satellite mass until
this lifetime is equal to the system age.  Sample evolutionary histories of this
maximum mass satellite for a hypothetical $1M_\odot$, $1M_{Jup}$ system from each
orbital evolutionary history category are shown in Figure \ref{figure:avst}.  The
numerically determined maximum mass as a function of planetary orbital semimajor
axis is shown in Figure \ref{figure:Mvsa} as the solid line.   These numerical
results are consistent with our analytical upper mass limits from Equations
\ref{eq:Rhilltransverse} and \ref{eq:Rhilltransverse2}.

\section{IMPLICATIONS}

\subsection{Known Extrasolar Planets}

In applying these results to the specific test case of the transiting planet
HD209458b, we adopt the values $M_*=1.1~\mathrm{M_\odot}$,
$M_p=0.69~\mathrm{M_{Jup}}$, $a_p=0.0468~\mathrm{AU}$, $T=5.0~\mathrm{Gyr}$
\citep{2000ApJ...532L..55M}, and $R_p=1.35~\mathrm{R_{Jup}}$
\citep{2001ApJ...552..699B} based on observational studies.  We take $k_{2p}$ for
the planet to be $0.51$, the value for an $n=1$ polytrope
\citep{1984QB603.I53H83...}.  The least constrained parameter is the tidal
dissipation factor $Q_p$; for HD209458b we adopt $Q_p=10^5$, which is consistent
with estimates for Jupiter's $Q_p$ \citep{1966Icar....5..375G}.  However,
$Q_p$ is not known precisely even for the planets in our own solar system, and the
precise mechanism for the dissipation of tidal energy has not been established. 
$Q_p$ for extrasolar planets, and especially for ones whose interiors differ from
Jupiter's such as close-in giant planets \citep{2000ApJ...534L..97B}, may differ
substantially from this value.  

Because HD209458b was likely tidally spun down to synchronous rotation very
quickly \citep{1996ApJ...459L..35G}, satellites around it are constrained by
the infall time and we use Equation \ref{eq:Rhilltransverse} to obtain an upper
limit of $7\times 10^{-7}~\mathrm{M_\oplus}$ for their masses. 
Assuming a density of $\rho = 3~\mathrm{g/cm^3}$, the largest possible
satellite would be $70~\mathrm{km}$ in radius --- slightly smaller than
Jupiter's irregularly-shaped moon Amalthea.  These limits are consistent with
the those placed on actual satellites observationally by
\citet{2001ApJ...552..699B}.  It is possible for captured satellites to exist
around HD209458b.  Their lifetimes, however, would be exceedingly short --- a
$1\mathrm{M_\oplus}$ satellite could survive for only $30,000\mathrm{yr}$
(Equation \ref{eq:retrotransverse}), making the probability of detecting one
low unless such captures are common.

We also calculate the maximum masses for surviving moons around other detected
extrasolar planets; the results are in Table \ref{table:lifetimes}.  We take the
planet masses to be equal to the minimum mass determined by radial velocity
monitoring, as the orbital inclination has not been reliably determined for any
planet except HD209458b.  We use the same $k_{2p}$ and $Q_p$ as we did for
HD209458b, but take $R_p=R_{Jup}$ because the radii for these objects is unknown. 
For this table, we have only chosen planets whose orbital eccentricities are less
than $0.10$ because Equation \ref{eq:acrit} applies only to planets in circular
orbits.  The critical semimajor axis for planets in noncircular orbits has not yet
been determined, thus we leave the calculation of upper mass limits for satellites
around these eccentric planets for future work.

\subsection{Earth-like Moons}

\begin{figure} 
\plotone{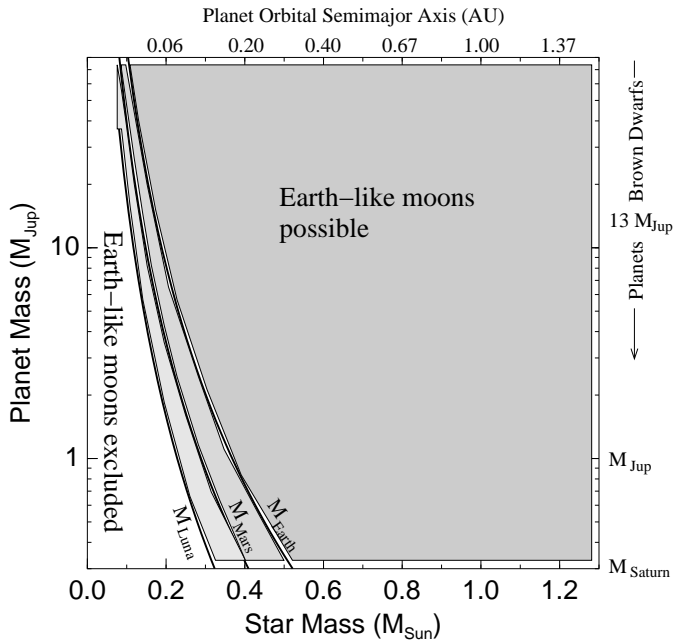}
\caption{Stability of Earth-like moons.  For stars of a given mass, we estimate a
luminosity from Equation \ref{eq:starLum}, and then calculate the limits on
planet masses that allow for the
stability of moons with insolation $F = 1370~\mathrm{W~m^{-2}}$ and 
mass $1~\mathrm{M_{Luna}}$ (solid line, left), 
$1~\mathrm{M_{Mars}}$ (solid line, center), and 
$1~\mathrm{M_{Earth}}$ (solid line, right).  Planets with the proper $a_p$ (top) and
masses above the appropriate solid line might still harbor Earth-like satellites
after $5~\mathrm{Gyr}$.  For this plot, we have taken $R_p = R_{Jup}$, 
and as such the values are not valid for planets with
masses less than $0.3~M_{Jup}$.  Planet mass limits for very low mass stars where 
$M_p \sim M_*$ are also suspect.  From this plot we infer that Earth-like
satellites of Jovian planets are plausibly stable for 
$5~\mathrm{Gyr}$ around most stars with masses greater than $0.15~M_\odot$.
\label{figure:habitable}} 
\end{figure}

Our approach can also shed light on the issue of Earth-like satellites, which we
define to be moons capable of supporting liquid water.  Low-mass satellites do not
fit this definition due to their inability to retain volatiles
\citep{1997Natur.385..234W}.  Here we note that high-mass satellites may not survive
for long periods around close-in planets because planets with low masses have
smaller Hill spheres and therefore, for a given satellite mass, also have shorter
maximum moon lifetimes.  To calculate in general which giant planets might harbor
Earth-like satellites, we use Equation \ref{eq:Rhilltransverse} and constrain $a_p$
based on the insolation at the planet, $F$, relative to the Earth's insolation
$F_\oplus=1370~\mathrm{W~m^{-2}}$.  We use the rough approximation 
\citep{1994sipp.book.....H}
\begin{equation}
\label{eq:starLum}
\frac{L_*}{L_\odot} = \left( \frac{M_*}{M_\odot} \right)^{3.5}
\end{equation}
for the stellar luminosity, $L_*$, together with the insolation at the planet,
\begin{equation}
\label{eq:flux}
\frac{F}{F_\oplus} \frac{a_p^2}{a_\oplus^2} = \frac{L_*}{L_\odot}~\mathrm{,}
\end{equation}
to fix the planet's semimajor axis by solving for $a_p$.  By plugging the resulting
value for $a_p$ into Equation \ref{eq:Rhilltransverse}, we can exclude Earth-like 
moons around planets in systems that don't satisfy the inequality
\begin{equation}
\label{eq:habitable}
M_p \ge \left[\frac{39}{4}\left(\frac{f^3}{3}\right)^{-13/6}
\frac{M_m k_{2p} T R_p^5 \sqrt{G}}{Q_p}
\right]^{3/8}
\frac{M_\odot^{273/64}}{M_*^{221/64}}
\left(a_\oplus^2\frac{F_\oplus}{F}\right)^{-39/32}~\mathrm{.}
\end{equation}
Equation
\ref{eq:habitable} is plotted in Figure \ref{figure:habitable} for the same values
of $k_{2p}$, $Q_p$, and $f$ as we use for HD209458b, with $R_p=R_{Jup}$ and
$T=5~\mathrm{Gyr}$.   

\citet{1997Natur.385..234W} found the lower limit $M_m \ge 0.12~\mathrm{M_\oplus}$
for moons that can retain volatiles over Gyr timescales.  Using this mass, we find
that Earth-like moons orbiting Jovian planets could survive for solar system
lifetimes around stars with masses greater than $0.15~M_\odot$, and that Earth-mass
satellites which receive similar insolation to the Earth are stable around all
Jovian planets orbiting stars with $M_* > 0.5~M_\odot$.  Planets with masses less
than $0.3~M_{Jup}$ differ in radius, $Q_p$, and interior structure from those with
masses greater than $0.3~M_{Jup}$.  In addition, for lower planet masses the
planet/satellite mass ratio increases beyond the assumption of non-synchronization
between the planet and moon.  For these reasons, we do not treat the question of
Earth-like satellites of ice-giant planets ($M_p<0.3\mathrm{M_{Jup}}$) here.

The radial-velocity planet most likely to harbor Earth-like moons is HD28185b
because of its circular, $a_p=1\mathrm{AU}$ orbit around a star similar to the
Sun withspectral type G5V and $L_*=1.09\mathrm{L_\odot}$
\citep{2001AA...379..999S}.  Because the calculated upper satellite mass for
this planet is above $8~\mathrm{M_\oplus}$, we can not rule out any satellite
masses for this object.  Thus, Earth-like moons with any mass could plausibly
be stable around HD28185b.

\subsection{Future Discoveries}

Several missions to search for extrasolar planet transits by high-precision
space-based photometry are in the planning stages and will, if launched, have the
capability of detecting satellites \citep{1999A&AS..134..553S}.  The probability
that a given planet will transit across its parent star decreases with planetary
orbital semimajor axis as $1/a_p$.  Hence these surveys will preferentially detect
planets in orbits close to their parent stars.  However, we have shown that it is
unlikely that these close-in objects will harbor satellites.  Therefore satellite
transits are most likely to be detected around planets orbiting at moderate
distances from their parent star ( $0.3 \mathrm{AU} \le a_p \le 2 \mathrm{AU}$),
even though planet transits are most likely at small orbital distances.  If a
satellite were detected, Equation \ref{eq:Rhilltransverse2} could be used to place
limits on the planetary tidal dissipation parameter $Q_p$.  By using extreme values
of the lifetimes, masses, and possible values $Q_p$ that may exist, we estimate this
process will not significantly affect planets more than $0.6\mathrm{AU}$ from their
parent star, leaving any satellite systems they might posess intact.

\acknowledgements

The authors acknowledge Bill Hubbard and Rick Greenberg for useful discussions,
and Robert H. Brown, Jonathan Fortney, Gwen Bart, and Paul Withers for manuscript
suggestions.  Funds for publication of this paper were provided by the Lunar and
Planetary Laboratory.  D. O. is supported by NASA GSRP.

\newpage

\bibliographystyle{apj}

\begin{deluxetable}{l|ll}
\tablecaption{Satellite Semimajor Axes\label{table:hillsemimajoraxes}}
\tablewidth{0pt}
\tablehead{
\colhead{Planet} &
\colhead{Satellite} & 
\colhead{${a_m}/{R_H}$}
}
\startdata
Earth  &	Moon			& 0.257 \\
Mars	 &	Deimos		& 0.0216\\
Jupiter& Callisto		& 0.0354\\
Jupiter& Elara			& 0.221 \\
Jupiter& Sinope		& 0.446\tablenotemark{R}\\
Saturn & Titan			& 0.0187\\
Saturn & Iapetus		& 0.0545\\
Saturn & Phoebe		& 0.198\tablenotemark{R}\\
Saturn & S/2000 S 9  & 0.283\\
Uranus & Oberon		& 0.00837\\
Uranus & Setebos		& 0.352\tablenotemark{R}\\
Neptune& Triton		& 0.003\tablenotemark{R}\\
Neptune& Nereid		& 0.0475\\
\enddata
\tablenotetext{R}{Retrograde}
\tablecomments{
Orbital semimajor axes of selected solar system satellites are listed as a 
function of their their parent planet's hill sphere radius, $R_H$.} 
\end{deluxetable}

\begin{deluxetable}{lccccccccc}
\tablecaption{Constraints on Satellites around Selected Extrasolar Planets\label{table:lifetimes}}
\tablewidth{0pt}
\tablehead{
\colhead{Name} &
\colhead{Star} &
\colhead{M sin i} &
\colhead{a} &
\colhead{e} &
\colhead{Max} &
\colhead{Max} &
\colhead{Reference} \\ 
\colhead{} &
\colhead{Age} &
\colhead{($M_{Jup}$)} &
\colhead{(AU)} &
\colhead{} &
\colhead{Moon} &
\colhead{Moon} \\
\colhead{} &
\colhead{(Gyr)} &
\colhead{} &
\colhead{} &
\colhead{} &
\colhead{Mass} &
\colhead{Radius} \\
\colhead{} &
\colhead{} &
\colhead{} &
\colhead{} &
\colhead{} &
\colhead{($M_{\oplus}$)} &
\colhead{(km)}
}

\startdata
     HD83443b & (5) & 0.35 & 0.038 &  0.08   & $ 8\times 10^{-8}$ & 30 & \citet{unpublished83443}\\
     HD46375b & (5) & 0.25 & 0.041 &  0.04   & $ 6\times 10^{-8}$ & 30 & \citet{2000ApJ...536L..43M}\\
    HD187123b & (5) & 0.52 & 0.042 &  0.03   & $ 6\times 10^{-7}$ & 60 & \citet{1998PASP..110.1389B}\\
    HD209458b &  5  & 0.69 & 0.045 &     0   & $ 7\times 10^{-7}$ & 70 & \citet{2002ApJ...564.1028F}\\
    HD179949b & (5) & 0.84 & 0.045 &     0   & $ 3\times 10^{-6}$ & 110 & \citet{2001ApJ...551..507T}\\
     HD75289b & (5) & 0.42 & 0.046 & 0.053   & $ 6\times 10^{-7}$ & 60 & \citet{2000AA...356..590U}\\
BD -10 3166 b & (5) & 0.48 & 0.046 &  0.05   & $ 7\times 10^{-7}$ & 70 & \citet{2000ApJ...545..504B}\\
      T Boo b &   2 &  4.1 & 0.047 & 0.051   & $ 5\times 10^{-4}$ & 600 & \citet{1997ApJ...474L.115B}\\
  51Pegasus b & (5) & 0.44 & 0.051 & 0.013   & $ 2\times 10^{-6}$ & 90 & \citet{1997ApJ...481..926M}\\
      U And b & 2.6 & 0.71 & 0.059 & 0.034   & $ 2\times 10^{-5}$ & 190 & \citet{1999ApJ...526..916B}\\
    HD168746b & (5) & 0.24 & 0.066 &     0   & $ 2\times 10^{-6}$ & 100 & \tablenotemark{1}\\
    HD130322b & (5) &    1 & 0.088 & 0.044   & $0.0008$ & 730 & \citet{2000AA...356..590U}\\
      55Cnc b &   5 & 0.84 &  0.11 & 0.051   & $ 0.001$ & 810 & \citet{1997ApJ...474L.115B}\\
        Gl86b & (5) &  3.6 &  0.11 & 0.042   & $   0.1$ & 3950 & \citet{2000AA...354...99Q}\\
    HD195019b & 3.2 &  3.5 &  0.14 &  0.03   & $   0.8$ & 7090 & \citet{1999PASP..111...50F}\\
       GJ876c &   5 &  1.9 &  0.21 &   0.1   & $     6$ & 14050 & \citet{2001ApJ...556..296M}\\
    rho CrB b &  10 &  1.1 &  0.23 & 0.028   & $   0.3$ & 5310 & \citet{1997ApJ...483L.111N}\\
      U And c & 2.6 &  2.1 &  0.83 & 0.018   &      \nodata & \nodata & \citet{1999ApJ...526..916B}\\
     HD28185b & (5) &  5.6 &     1 &  0.06   &      \nodata & \nodata & \citet{2001AA...379..999S}\\
     HD27442b & (5) &  1.4 &   1.2 &  0.02   &      \nodata & \nodata & \citet{2001ApJ...555..410B}\\
    HD114783b & (5) &    1 &   1.2 &   0.1   &      \nodata & \nodata & \citet{2002ApJ...568..352V}\\
     HD23079b & (5) &  2.5 &   1.5 &  0.02   &      \nodata & \nodata & \citet{2001astro.ph.11255T}\\
      HD4208b & (5) &  0.8 &   1.7 &  0.01   &      \nodata & \nodata & \citet{2002ApJ...568..352V}\\
      47UMa b & 6.9 &  2.5 &   2.1 & 0.061   &      \nodata & \nodata & \citet{2002ApJ...564.1028F}\\
      47UMa c & 6.9 & 0.76 &   3.7 &   0.1   &      \nodata & \nodata & \citet{2002ApJ...564.1028F}\\
\enddata
\tablenotetext{1}{http://obswww.unige.ch/\~~\hspace{-4pt}udry/planet/hd168746.html}
\tablenotetext{2}{http://c3po.lpl.arizona.edu/egpdb}
\tablecomments{
Upper satellite mass limits are determined with $Q_p=10^5$, $R_p=R_{Jup}$,
and with $M_p=m\sin i$.  Where no system lifetime was available in the literature, we have
taken the system age to be $5~\mathrm{Gyr}$, and those cases are indicated by
parenthesis.  To obtain maximum moon radii, we assume $\rho=3~
\mathrm{g~cm^{-3}}$.  We can not place useful limits for those planets where 
maximum masses and radii are not listed.  Planets with
orbital eccentricities greater than $0.1$ are excluded due to the difficulty
in determining the proper value of $f$.  
The planet data used to generate this table have been formed 
into a
world wide web accessible database of extrasolar planets
\tablenotemark{2}.   }
\end{deluxetable}

\end{document}